\documentstyle[prl,aps,preprint]{revtex}
\begin{document}

\bigskip\ 

\bigskip\ 

\begin{center}
{\bf REMARKS ON WEYL INVARIANT P-BRANES AND Dp-BRANES}

{\bf \ }

{\bf \ }

\smallskip\ 

J. A. Nieto\footnote{%
nieto@uas.uasnet.mx}

\smallskip

{\it Facultad de Ciencias F\'{\i}sico-Matem\'{a}ticas de la Universidad
Aut\'{o}noma}

{\it de Sinaloa, C.P. 80010, Culiac\'{a}n Sinaloa, M\'{e}xico}

\bigskip\ 

\bigskip\ 

{\bf Abstract}
\end{center}

A mechanism to find different Weyl invariant p-branes and Dp-branes actions
is explained. Our procedure clarifies the Weyl invariance for such systems.
Besides, by considering gravity-dilaton effective action in higher dimensions
we also derive a Weyl invariant action for p-branes. We argue that this
derivation provides a geometrical scenario for the Weyl invariance of
p-branes. Our considerations can be extended to the case of super-p-branes.

\smallskip\ 

Pacs numbers: 11.10.Kk, 04.50.+h, 12.60.-i

October/2001

\newpage\ \noindent {\bf 1.- INTRODUCTION}

\bigskip

It is known that, in string theory, the Weyl invariance of the Polyakov
action plays a central role [1]. In fact, in string theory subjects such as
moduli space, Teichmuler space and critical dimensions, among many others,
are consequences of the local diff x Weyl symmetry in the partition function
associated to the Polyakov action.

In the early eighties there was a general believe that the Weyl invariance
was the key symmetry to distinguish string theory from other p-branes.
However, in 1986 it was noticed that such an invariance may be also
implemented to any p-brane and in particular to the 3-brane [2]. The formal
relation between the Weyl invariance and p-branes was established two years
later independently by a number of authors [3]. It spites of its possible
relevance in M-theory, it seems that except for some works [4] the subject
has been very much ignored. Recently, however, the importance of the Weyl
invariance has been revived in connection with 2d-gravity [5] and Dp-branes
[6].

On the other hand, in a recent work [7] it was proved that it is possible to
obtain the p-brane action from a gravity-dilaton effective action in higher
dimensions. This work has some relevance, among other things, because it
opens the possibility to understand symmetries of the p-branes from a
geometrical point of view. In particular, it may be of special interest to
understand the origin of the Weyl invariance for p-branes from a geometrical
scenario. In this article, we prove that using a particular ansatz for the
metric, a gravity-dilaton action in $D+1$ dimensions is reduced to a Weyl
invariant p-brane action, clarifying with this the geometric origin of Weyl
invariance for p-branes. As reference [7], the present work may be of
special interest in the Randall-Sundrum scenario [8-9], string theory [10]
and M-theory [11].

We start making some general remarks on the Weyl invariant p-branes and
D-branes. Specifically, we show a mechanism to obtain the Weyl invariant
actions for p-branes and D-branes from the Dirac-Nambu-Goto p-brane action
and the Born-Infeld Dp-brane action, respectively. We also apply a similar
mechanism to the case of null p-branes and derive a Weyl invariant action
for such system.

The plan of this work is as follows. In section 2, we show a mechanism to
obtain Weyl invariant p-branes from the Dirac-Nambu-Goto p-brane action. In
section 3, we apply a similar procedure in order to obtain Weyl invariant
Dp-branes from the Born-Infeld Dp-brane action. In section 4, we derive a
Weyl invariant action for null p-branes. In section 5, we obtain a Weyl
invariant p-brane action from a gravity-dilaton effective action. Finally, in
section 6 we make some final comments.

\smallskip\ 

\noindent {\bf 2.- WEYL INVARIANT P-BRANES FROM DIRAC-NAMBU-GOTO \ }

\ {\bf P-BRANE ACTION}

\bigskip

Consider a p-brane moving in a $d+1$-dimensional Minkowski space-time. The
evolution of such a system may be described by using the $d+1$-scalar field
coordinates $x^{\hat{\mu}}(\xi ^{A})$ where $\hat{\mu}=0,1,...,d$ and $\xi
^{A}$, with $A=0,1,...,p,$ are arbitrary parameters.

The Dirac-Nambu-Goto type action for p-branes is

\begin{equation}
S_{p}^{(1)}=-T_{p}\int d^{p+1}\xi \sqrt{-h},  \eqnum{1}
\end{equation}
where $h=\det (h_{AB}),$ with

\begin{equation}
h_{AB}=\partial _{A}x^{\hat{\mu}}\partial _{B}x^{\hat{\nu}}\eta _{\hat{\mu}%
\hat{\nu}},  \eqnum{2}
\end{equation}
and $T_{p}$ is a fundamental constant measuring the inertia of the p-brane.
Here,

\begin{equation}
\eta _{\hat{\mu}\hat{\nu}}=diag(-1,1,...,1)  \eqnum{3}
\end{equation}
is the Minkowski metric.

Let us now consider a symmetric auxiliary world-volume metric $%
g_{AB}=g_{AB}(\xi ^{A})$ and let us assume that the integrand $\sqrt{-h}$ in
(1) is replaced by $\sqrt{-g}$ where $g=\det (g_{AB})$. The question is how
to get an equivalent action to (1) with $\sqrt{-g}$ as the integrand instead
of $\sqrt{-h}.$ Clearly, the integrands $\sqrt{-g}$ and $\sqrt{-h}$ are
equivalent if we assume the constraint

\begin{equation}
h_{AB}-g_{AB}=0.  \eqnum{4}
\end{equation}
So, by introducing the lagrange multipliers $\Lambda ^{AB}=\Lambda ^{BA}$ we
discover the action

\begin{equation}
S_{p}^{(2)}=-T_{p}\int d^{p+1}\xi (\sqrt{-g}+\Lambda ^{AB}(h_{AB}-g_{AB})) 
\eqnum{5}
\end{equation}
which is equivalent to $S_{p}^{(1)}$. In fact, the $\Lambda ^{AB}$ field
equation implies the constraint (4) which allows to reduce (5) to (1). The
point is that $g_{AB}$ and $\Lambda ^{AB}$ provide to the action $%
S_{p}^{(2)} $ with more degrees of freedom than $S_{p}^{(1)}$, but these
additional degrees of freedom can be eliminated by means of (4) reducing $%
S_{p}^{(2)}$ to $S_{p}^{(1)}$.

We may obtain another equivalent action to $S_p^{(1)}$ if instead of varying 
$\Lambda ^{AB}$ in (5) we make variations of $g_{AB}.$ We find that $g_{AB}$
field equation is

\begin{equation}
\frac 12\sqrt{-g}g^{AB}-\Lambda ^{AB}=0.  \eqnum{6}
\end{equation}
This equation allows to eliminate $\Lambda ^{AB}$ from (5). Thus, we obtain
the alternative action

\begin{equation}
S_{p}^{(3)}=-\frac{T_{p}}{2}\int d^{p+1}\xi \sqrt{-g}(g^{AB}h_{AB}-(p-1)) 
\eqnum{7}
\end{equation}
which is the familiar Polyakov type action for p-branes. Thus, starting from
the Dirac-Nambu-Goto action (1) we have found a mechanism to derive the
Polyakov action for p-branes (7). The procedure to derive $S_{p}^{(1)}$
starting from $S_{p}^{(3)}$ is well known. One first derive from $%
S_{p}^{(3)} $ the $g^{AB}$ field equation

\begin{equation}
h_{AB}-\frac{1}{2}g_{AB}(g^{CD}h_{CD}-(p-1))=0.  \eqnum{8}
\end{equation}
Multiplying this expression by $g^{AB},$ it is not difficult to obtain, for $%
p\neq 1,$ the formula

\begin{equation}
g^{AB}h_{AB}=p+1.  \eqnum{9}
\end{equation}
Substituting (9) into (8) leads us to the constraint (4) which allows to
reduce (7) to (1).

Let us now follow a slightly different procedure. Consider the rescaling of
the metric $g_{AB}\rightarrow \Phi g_{AB}$, where $\Phi =\Phi (\xi )$ is a
scale factor. If we apply this transformation to (5), leaving $\Lambda ^{AB}$
and $h_{AB}$ unchanged, we obtain

\begin{equation}
S_{p}^{(\hat{2})}=-T_{p}\int d^{p+1}\xi (\Phi ^{\frac{p+1}{2}}\sqrt{-g}%
+\Lambda ^{AB}(h_{AB}-\Phi g_{AB})).  \eqnum{10}
\end{equation}
Varing $g_{AB}$ in (10) leads to

\begin{equation}
\frac{1}{2}\sqrt{-g}\Phi ^{\frac{p+1}{2}}g^{AB}-\Phi \Lambda ^{AB}=0. 
\eqnum{11}
\end{equation}
Solving for $\Lambda ^{AB}$ gives

\begin{equation}
\Lambda ^{AB}=\frac{1}{2}\sqrt{-g}\Phi ^{\frac{p-1}{2}}g^{AB}.  \eqnum{12}
\end{equation}
Substituting this result back into (10) yields

\begin{equation}
S_{p}^{W_{1}}=-\frac{T_{p}}{2}\int d^{p+1}\xi \sqrt{-g}(\Phi ^{\frac{p-1}{2}%
}g^{AB}h_{AB}-(p-1)\Phi ^{\frac{p+1}{2}}).  \eqnum{13}
\end{equation}
Note that if we set $\Phi =1$ then (13) is reduced to (7). Therefore, the
action $S_{p}^{W_{1}}$ is a generelization of $S_{p}^{(3)}$. We also
discover that (13) is invariant under the transformations

\begin{equation}
\Phi \rightarrow \lambda ^{-1}\Phi  \eqnum{14}
\end{equation}
and

\begin{equation}
g_{AB}\rightarrow \lambda g_{AB}.  \eqnum{15}
\end{equation}
where $\lambda (\xi )$ is an arbitrary function. Of course, (15) is a Weyl
transformation. In order to clarify this point let us integrate out $\Phi $
from (13). We find the formula

\begin{equation}
\Phi =\frac{g^{AB}h_{AB}}{p+1}.  \eqnum{16}
\end{equation}
Substituting this result into (13) leads to

\begin{equation}
S_{p}^{(4)}=-\frac{T_{p}}{(p+1)^{\frac{p+1}{2}}}\int d^{p+1}\xi \sqrt{-g}%
(g^{AB}h_{AB})^{\frac{p+1}{2}}.  \eqnum{17}
\end{equation}
We recognize this action as the Weyl invariant action for p-branes [2]-[3].
In fact, the action $S_{p}^{(4)}$ is invariant under the Weyl transformation
(15). Moreover, it is straightforward to see that the Weyl invariant action $%
S_{p}^{(1)}$ is recovered upon eliminating $g_{AB\text{ }}$from $S_{p}^{(4)}$
via its field equation, showing with this that $S_{p}^{(4)}$ is classically
equivalent to $S_{p}^{(1)}$.

If we redefine $\Phi $ as

\begin{equation}
\Phi =\phi ^{\frac{2}{p+1}}  \eqnum{18}
\end{equation}
we find that (13) becomes

\begin{equation}
S_{p}^{W_{2}}=-\frac{T_{p}}{2}\int d^{p+1}\xi \sqrt{-g}(\phi ^{\frac{p-1}{p+1%
}}g^{AB}h_{AB}-(p-1)\phi )  \eqnum{19}
\end{equation}
which is the action proposed in reference [6].

An interesting possibility arises if instead of using (18) we redefine $\Phi 
$ as

\begin{equation}
\Phi =\varphi ^{\frac{2}{p-1}},  \eqnum{20}
\end{equation}
with $p-1\neq 0$. In fact, in this case the action (13) becomes

\begin{equation}
S_{p}^{W_{3}}=-\frac{T_{p}}{2}\int d^{p+1}\xi \sqrt{-g}(\varphi
g^{AB}h_{AB}-(p-1)\varphi ^{\frac{p+1}{p-1}}).  \eqnum{21}
\end{equation}
This action is interesting because in the second term of (21) the scalar
field $\varphi $ gives a power potential. The case $p=1$ is solved if in
(21) we set $\varphi =1.$

\smallskip\ \smallskip\ 

\noindent {\bf 3.- WEYL INVARIANT Dp-BRANES FROM BORN-INFELD Dp-BRANE }

\ \ {\bf ACTION}

\bigskip

In this section, we shall show that a similar procedure, as in the previous
section, can be applied to the case of Dp-branes. Consider the Born-Infeld
action for Dp-branes

\begin{equation}
S_{Dp}^{(1)}=-T_{p}\int d^{p+1}\xi \sqrt{-f},  \eqnum{22}
\end{equation}
where $T_{p}$ is the Dp-brane ``tension'' and $f=\det (f_{AB})$ with

\[
f_{AB}=h_{AB}-{\cal F}_{AB}. 
\]
Here, ${\cal F}_{AB}$ is a two form given by

\begin{equation}
{\cal F}_{AB}=F_{AB}-B_{AB},  \eqnum{23}
\end{equation}
where $F_{AB}$ is the field strength

\begin{equation}
F_{AB}=\partial _{A}A_{B}-\partial _{B}A_{A}  \eqnum{24}
\end{equation}
of a U(1) gauge field $A_{B}$ and $B_{AB}$ is the NS antisymmetric two form
field. Note that if ${\cal F}_{AB}=0$ then (22) is reduced to (1) and
therefore (22) is a generalization of the Dirac-Nambu-Goto type action for
p-branes.

Following a similar procedure as in the previous section we introduce an
auxiliary metric world-volume metric $s_{AB}=s_{AB}(\xi ^{A}).$ In contrast
to $g_{AB}$ the second rank tensor $s_{AB}$ has no symmetries. If we
consider the constraint

\begin{equation}
f_{AB}-s_{AB}=0  \eqnum{25}
\end{equation}
and introduce the lagrange multipliers $\Sigma ^{AB},$ it is straightforward
to show that the action

\begin{equation}
S_{Dp}^{(2)}=-T_{p}\int d^{p+1}\xi (\sqrt{-s}+\Sigma ^{AB}(f_{AB}-s_{AB})) 
\eqnum{26}
\end{equation}
is equivalent to $S_{Dp}^{(1)}$.

Varying $s_{AB}$ in $S_{Dp}^{(2)}$ we find the formula

\begin{equation}
\frac 12\sqrt{-s}s^{AB}-\Sigma ^{AB}=0.  \eqnum{27}
\end{equation}
This equation allows to eliminate $\Sigma ^{AB}$ from (26). We obtain

\begin{equation}
S_{Dp}^{(3)}=-\frac{T_{p}}{2}\int d^{p+1}\xi \sqrt{-s}(s^{AB}f_{AB}-(p-1)) 
\eqnum{28}
\end{equation}
which is the analogue of the Polyakov type action for p-branes (see Ref.
[12]). Using the field equation for $s_{AB}$ one can recover the action $%
S_{Dp}^{(1)}$ from $S_{Dp}^{(3)}$. So, $S_{Dp}^{(1)}$ and $S_{Dp}^{(3)}$ are
classically equivalent actions.

Now, consider the rescaling $s_{AB}\rightarrow \Phi s_{AB}$, where $\Phi
(\xi )$ is again a scale factor function and apply this transformation to
(26), leaving $\Sigma ^{AB}$ and $f_{AB}$ unchanged. We obtain

\begin{equation}
S_{Dp}^{(\hat{2})}=-T_p\int d^{p+1}\xi (\Phi ^{\frac{p+1}2}\sqrt{-s}+\Sigma
^{AB}(f_{AB}-\Phi s_{AB})).  \eqnum{29}
\end{equation}
Varying $s_{AB}$ we find

\begin{equation}
\frac{1}{2}\sqrt{-g}\Phi ^{\frac{p+1}{2}}s^{AB}-\Phi \Sigma ^{AB}=0. 
\eqnum{30}
\end{equation}
Solving for $\Lambda ^{AB}$ and substituting the result back into (29) yields

\begin{equation}
S_{Dp}^{W_{1}}=-\frac{T_{p}}{2}\int d^{p+1}\xi \sqrt{-s}(\Phi ^{\frac{p-1}{2}%
}s^{AB}f_{AB}-(p-1)\Phi ^{\frac{p+1}{2}}).  \eqnum{31}
\end{equation}
We discover that $S_{Dp}^{W_{1}}$ is invariant under the transformations

\begin{equation}
\Phi \rightarrow \lambda ^{-1}\Phi  \eqnum{32}
\end{equation}
and

\begin{equation}
s_{AB}\rightarrow \lambda s_{AB}.  \eqnum{33}
\end{equation}

Integrating $\Phi $ in (31) we find the field equation

\begin{equation}
\Phi =\frac{s^{AB}f_{AB}}{p+1}.  \eqnum{34}
\end{equation}
Susbtituting this result into (31) leads to

\begin{equation}
S_{Dp}^{(4)}=-\frac{T_{p}}{(p+1)^{\frac{p+1}{2}}}\int d^{p+1}\xi \sqrt{-s}%
(s^{AB}f_{AB})^{\frac{p+1}{2}}  \eqnum{35}
\end{equation}
which is a generalization of the Weyl invariant action for p-branes (17)
(see Ref. [13]).

If we redefine $\Phi $ as

\begin{equation}
\Phi =\phi ^{\frac{2}{p+1}},  \eqnum{36}
\end{equation}
we find that (31) becomes

\begin{equation}
S_{Dp}^{W_{2}}=-\frac{T_{p}}{2}\int d^{p+1}\xi \sqrt{-s}(\phi ^{\frac{p-1}{%
p+1}}s^{AB}f_{AB}-(p-1)\phi ).  \eqnum{37}
\end{equation}

Similar, if we redefine $\Phi $ as

\begin{equation}
\Phi =\varphi ^{\frac{2}{p-1}},  \eqnum{38}
\end{equation}
with $p-1\neq 0,$ we find

\begin{equation}
S_{Dp}^{W_{3}}=-\frac{T_{p}}{2}\int d^{p+1}\xi \sqrt{-s}(\varphi
s^{AB}f_{AB}-(p-1)\varphi ^{\frac{p+1}{p-1}}).  \eqnum{39}
\end{equation}
It seems that the actions $S_{Dp}^{W_{1}},S_{Dp}^{W_{2}}$ or $S_{Dp}^{W_{3}}$
have not been considered in the literature. In reference [6] a Weyl
invariant action for Dp-branes was proposed, but the authors used the action
of Zeid and Hull [14] rather than $S_{Dp}^{(3)}$ action.

\smallskip\ \smallskip\ 

\noindent {\bf 4.- WEYL INVARIANT NULL P-BRANES }

\bigskip

The procedure may also work out for null p-branes. Let us show in this
section this fact. Null p-branes are defined for the case $T_{p}=0.$ So, the
action (1) is not appropriate for this case. The key is first to write $h$ as

\begin{equation}
h=\sigma ^{\hat{\mu}_{1}...\hat{\mu}_{p+1}}\sigma _{\hat{\mu}_{1}...\hat{\mu}%
_{p+1}},  \eqnum{40}
\end{equation}
where

\begin{equation}
\sigma ^{\hat{\mu}_{1}...\hat{\mu}_{p+1}}=\frac{1}{[(p+1)!]^{\frac{1}{2}}}%
\varepsilon ^{A_{1}...A_{p+1}}\partial _{A_{1}}x^{\hat{\mu}_{1}}...\partial
_{A_{p+1}}x^{\hat{\mu}_{p+1}}.  \eqnum{41}
\end{equation}
Then, one can show that the action (1) is equivalent to

\begin{equation}
S_{p}^{(1)}=\int d^{p+1}\xi (\sigma ^{\hat{\mu}_{1}...\hat{\mu}_{p+1}}p_{%
\hat{\mu}_{1}...\hat{\mu}_{p+1}}-\frac{\gamma }{2}(p^{\hat{\mu}_{1}...\hat{%
\mu}_{p+1}}p_{\hat{\mu}_{1}...\hat{\mu}_{p+1}}+T_{p}^{2})).  \eqnum{42}
\end{equation}
where $\gamma $ is a lagrange multiplier. If we eliminate $p_{\hat{\mu}%
_{1}...\hat{\mu}_{p+1}}$ from this action we get

\begin{equation}
S_{p}^{(1)}=\frac{1}{2}\int d^{p+1}\xi (\gamma ^{-1}\sigma ^{\hat{\mu}_{1}...%
\hat{\mu}_{p+1}}\sigma _{\hat{\mu}_{1}...\hat{\mu}_{p+1}}-\gamma T_{p}^{2}).
\eqnum{43}
\end{equation}
By eliminating $\gamma $ from (43) leads us to recover action (1). The
importance of (42) or (43) is that it now makes sense to set $T_{p}=0.$ In
this case (43) is reduced to the Schild type null p-brane action [15]

\begin{equation}
S_{Np}^{(1)}=\frac{1}{2}\int d^{p+1}\xi \gamma ^{-1}h.  \eqnum{44}
\end{equation}

Now, consider the equivalent action

\begin{equation}
S_{Np}^{(2)}=\frac{1}{2}\int d^{p+1}\xi (\gamma ^{-1}g+\Lambda
^{AB}(h_{AB}-g_{AB})).  \eqnum{45}
\end{equation}
The $g_{AB}$ field equation is

\begin{equation}
\gamma ^{-1}gg^{AB}-\Lambda ^{AB}=0.  \eqnum{46}
\end{equation}
Using this expression we learn that (45) is reduced to

\begin{equation}
S_{Np}^{(2)}=\frac{1}{2}\int d^{p+1}\xi \gamma ^{-1}g(g^{AB}h_{AB}-p). 
\eqnum{47}
\end{equation}

Considering the rescaling $g_{AB}\rightarrow \Phi g_{AB}$ we see that the
action (45) becomes

\begin{equation}
S_{Np}^{(\hat{2})}=\int d^{p+1}\xi (\Phi ^{p+1}\gamma ^{-1}g+\Lambda
^{AB}(h_{AB}-\Phi g_{AB})).  \eqnum{48}
\end{equation}
Now, we have

\begin{equation}
\gamma ^{-1}\Phi ^{p+1}gg^{AB}-\Phi \Lambda ^{AB}=0.  \eqnum{49}
\end{equation}
Therefore, we obtain

\begin{equation}
S_{Np}^{W_{1}}=\int d^{p+1}\xi (\gamma ^{-1}g(\Phi ^{p}g^{AB}h_{AB}-p\Phi
^{p+1})).  \eqnum{50}
\end{equation}

Two alternative actions are

\begin{equation}
S_{Np}^{W_{2}}=\int d^{p+1}\xi (\gamma ^{-1}g(\phi ^{\frac{p}{p+1}%
}g^{AB}h_{AB}-p\phi ))  \eqnum{51}
\end{equation}
and

\begin{equation}
S_{p}^{W_{3}}=\int d^{p+1}\xi (\gamma ^{-1}g(\varphi g^{AB}h_{AB}-p\varphi ^{%
\frac{p+1}{p}})),  \eqnum{52}
\end{equation}
depending if we redefine $\Phi $ as $\Phi =\phi ^{\frac{1}{p+1}}$ or $\Phi
=\varphi ^{\frac{1}{p}},$ respectively$.$

In (52) varying $\varphi $ we find the field equation

\begin{equation}
\varphi =\frac{1}{(p+1)^{p}}(g^{AB}h_{AB})^{p}.  \eqnum{53}
\end{equation}
Therefore, $S_{Np}^{W_{3}}$ becomes

\begin{equation}
S_{Np}^{(4)}=\frac{(1-p)}{(1+p)^{p}}\int d^{p+1}\xi \gamma
^{-1}g(g^{AB}h_{AB})^{p+1}  \eqnum{54}
\end{equation}
which is the Weyl invariant action for null p-branes [16].

\smallskip\ \smallskip\ 

\noindent {\bf 5.- WEYL INVARIANT P-BRANES FROM A GRAVITY-DILATON ACTION}

\bigskip

Here, we shall closely follow reference [7]. Our starting point is the
gravity-dilaton effective action with cosmological constant:

\begin{equation}
S=-%
%TCIMACRO{\tfrac{1}{16\pi G_{D+1}}}%
%BeginExpansion
{\textstyle{1 \over 16\pi G_{D+1}}}%
%EndExpansion
\int d^{D+1}y\sqrt{-g}e^{-\phi }(\varphi (R+(\nabla \phi )^{2})+2\Lambda
\varphi ^{a}),  \eqnum{55}
\end{equation}
where $G_{D+1\text{ }}$is the Newton constant in $D+1$ dimensions, $\phi
=\phi (y^{\alpha })$ is the dilaton field, $\Lambda $ and $a$ are constants, 
$\varphi $ is a lagrange multiplier and $R$ is the Ricci scalar obtained
from the Riemann tensor

\begin{equation}
R_{\nu \alpha \beta }^{\mu }=\Gamma _{\nu \beta ,\alpha }^{\mu }-\Gamma
_{\nu \alpha ,\beta }^{\mu }+\Gamma _{\sigma \alpha }^{\mu }\Gamma _{\nu
\beta }^{\sigma }-\Gamma _{\sigma \beta }^{\mu }\Gamma _{\nu \alpha
}^{\sigma }  \eqnum{56}
\end{equation}
and the metric tensor $g_{\alpha \beta }$, with $\alpha ,\beta =0,1...,D$.
Here, $\Gamma _{\alpha \beta }^{\mu }$ is the Christoffel symbol:

\begin{equation}
\Gamma _{\alpha \beta }^{\mu }=\frac{1}{2}g^{\mu \nu }(g_{\nu \alpha ,\beta
}+g_{\nu \beta ,\alpha }-g_{\alpha \beta ,\nu }).  \eqnum{57}
\end{equation}

Consider the ansatz

\begin{equation}
\begin{array}{ccc}
g_{AB} & = & \;\!\!\!\!\!\!\!\!\!\!\!\!\tilde{g}_{AB}(y^C), \\ 
&  &  \\ 
g_{ij} & = & a_k(y^C)a_l(y^C)\eta _{ij}^{kl}, \\ 
&  &  \\ 
g_{Ai} & = & \;\!\!\!\!\!\!\!\!\!\!\!\!\!\!\!\!\!\!\!\!\!\!\!0.
\end{array}
\eqnum{58}
\end{equation}
Here, the indices $A,B,...etc.$ run from $0$ to $p,$ the indices $%
i,j,...etc. $ run from $p+1$ to $D$ and the only non-vanishing terms of $%
\eta _{ij}^{kl}$ are

\begin{equation}
\eta _{ij}^{kl}=1\text{, when }k=l=i=j.  \eqnum{59}
\end{equation}

Assume that

\begin{equation}
\phi =\phi (y^{C})  \eqnum{60}
\end{equation}
and

\begin{equation}
\varphi (y^{C}).  \eqnum{61}
\end{equation}

Using (56), (57) and (58) one can compute the non-vanishing Chistoffel
symbols and the Reimann tensor. The result for the Ricci scalar\ $R=g^{\mu
\nu }R_{\mu \nu }$ (see Ref. [7] for detail computation, and Ref. [17] for
the case of $p=0$) is

\begin{equation}
\begin{array}{ccc}
R & = & -2a^iD^A\partial _Aa_i-a^ia^j\partial _Aa_i\partial ^Aa_j \\ 
&  &  \\ 
&  & \!\!\!\!\!\!\!\!+a^ia^j\partial _Aa_k\partial ^Aa_l\eta _{ij}^{kl}+%
\tilde{R},
\end{array}
\eqnum{62}
\end{equation}
where $\tilde{R}$ is the Ricci scalar associated to the metric $\tilde{g}%
_{AB}$.

Thus, the action (55) becomes

\begin{equation}
\begin{array}{ccc}
S & = & -\frac 1{16\pi G_{p+1}}\int d^{p+1}y\sqrt{-\tilde{g}}\Pi a_se^{-\phi
}\{\varphi (-2a^iD^A\partial _Aa_i-a^ia^j\partial ^Aa_i\partial _Aa_j \\ 
&  &  \\ 
&  & +a^ia^j\partial ^Aa_k\partial _Aa_l\eta _{ij}^{kl}+\partial ^A\phi
\partial _A\phi +\tilde{R})+2\Lambda \varphi ^a\},
\end{array}
\eqnum{63}
\end{equation}
where $G_{p+1\text{ }}$is the Newton constant in $p+1$ dimensions. The
relation between $G_{p+1\text{ }}$and $G_{D+1\text{ }}$is

\begin{equation}
\frac{1}{G_{p+1\text{ }}}=\frac{V_{d}}{G_{D+1\text{ }}},  \eqnum{64}
\end{equation}
where $V_{d}$ is a volume element in $d=D-p$ dimensions. This action can be
rewritten as

\begin{equation}
\begin{array}{c}
\!\!\!\!\!\!\!\!\!\!\!\!\!\!\!\!\!\!\!\!\!\!\!\!\!\!\!\!\!\!\!\!\!\!\!\!\!\!%
\!\!\!\!\!\!\!\!\!\!\!\!\!\!\!\!\!\!\!\!\!\!S=-\frac 1{16\pi G_{p+1}}\int
d^{p+1}y\sqrt{-\tilde{g}}D^AJ_A \\ 
\\ 
\!\!\!\!\!\!\!\!\!\!\!\!\!\!\!-\frac 1{16\pi G_{p+1}}\int d^{p+1}y\sqrt{-%
\tilde{g}}\Pi a_se^{-\phi }\{\varphi (a^ia^j\partial ^Aa_i\partial
_Aa_j-2\partial ^A\phi a^i\partial _Aa_i \\ 
\\ 
+\partial ^A\phi \partial _A\phi -a^ia^j\partial ^Aa_k\partial _Aa_l\eta
_{ij}^{kl})+2\Lambda \varphi ^a\} \\ 
\\ 
\!\!\!\!\!\!-\frac 1{16\pi G_{p+1}}\int d^{p+1}y\sqrt{-\tilde{g}}\Pi
a_se^{-\phi }\varphi \tilde{R}+\frac 1{16\pi G_{p+1}}\int d^{p+1}y\sqrt{-%
\tilde{g}}J_A\varphi ^{-1}\partial ^A\varphi ,
\end{array}
\eqnum{65}
\end{equation}
where

\begin{equation}
J_{A}=(-2\Pi a_{s}e^{-\phi }\varphi a^{i}\partial _{A}a_{i}).  \eqnum{66}
\end{equation}
Dropping the total derivative, the action (65) is reduced to

\begin{equation}
\begin{array}{c}
S=-\frac 1{16\pi G_{p+1}}\int d^{p+1}y\sqrt{-\tilde{g}}\Pi a_se^{-\phi
}\{\varphi (a^ia^j\partial ^Aa_i\partial _Aa_j-2\partial ^A\phi a^i\partial
_Aa_i+ \\ 
\\ 
+\partial ^A\phi \partial _A\phi -a^ia^j\partial ^Aa_k\partial _Aa_l\eta
_{ij}^{kl})+2\Lambda \varphi ^a\} \\ 
\\ 
\!-\frac 1{16\pi G_{p+1}}\int d^{p+1}y\sqrt{-\tilde{g}}\Pi a_se^{-\phi
}\varphi \tilde{R}+\frac 1{16\pi G_{p+1}}\int d^{p+1}y\sqrt{-\tilde{g}}%
J_A\varphi ^{-1}\partial ^A\varphi .
\end{array}
\eqnum{67}
\end{equation}

If we define the coordinate $x^{0}$ as

\begin{equation}
\Pi a_{s}e^{-\phi }=e^{-x^{0}},  \eqnum{68}
\end{equation}
the brane coupling ``constant'' $T_{p}$ in the form

\begin{equation}
\frac{e^{-x^{0}}}{16\pi G_{p+1}}=\frac{1}{2T_{p}}  \eqnum{69}
\end{equation}
and the variables $x^{i}$ as

\begin{equation}
x^{i}\equiv \ln a_{i}  \eqnum{70}
\end{equation}
we find\ that (67) becomes

\begin{equation}
\begin{array}{c}
S=\frac{1}{2}\int \frac{d^{p+1}y}{T_{p}}\sqrt{-\tilde{g}}\varphi (\tilde{g}%
^{AB}\partial _{A}x^{\hat{\mu}}\partial _{B}x^{\hat{\nu}}\eta _{\hat{\mu}%
\hat{\nu}})-2\Lambda \varphi ^{a}] \\ 
\\ 
-\frac{1}{2}\int \frac{d^{p+1}y}{T_{p}}\sqrt{-\tilde{g}}\varphi \tilde{R}+%
\frac{1}{16\pi G_{p+1}}\int d^{p+1}y\sqrt{-\tilde{g}}J_{A}\varphi
^{-1}\partial ^{A}\varphi .
\end{array}
\eqnum{71}
\end{equation}
where $\eta _{\hat{\mu}\hat{\nu}}=diag(-1,1,...,1).$ Here the indices $\hat{%
\mu},\hat{\nu},...$etc run from $0$ to $d=D-p.$

Let us assume now the case in which we can drop the last two terms from
(71). Setting the constant $\Lambda $ as

\begin{equation}
\Lambda =\frac{p-1}{2}  \eqnum{72}
\end{equation}
and the quantity $a$ as

\begin{equation}
a=\frac{p+1}{p-1}  \eqnum{73}
\end{equation}
we obtain

\begin{equation}
S=\frac{1}{2}\int \frac{d^{p+1}y}{T_{p}}\sqrt{-\tilde{g}}(\varphi \tilde{g}%
^{AB}h_{AB}-(p-1)\varphi ^{\frac{p+1}{p-1}}),  \eqnum{74}
\end{equation}
where 
\begin{equation}
h_{AB}=\partial _{A}x^{\hat{\mu}}\partial _{B}x^{\hat{\nu}}\eta _{\hat{\mu}%
\hat{\nu}}.  \eqnum{75}
\end{equation}
We recognize (74) as the action $S_{p}^{W_{3}}$ given in (21). Integrating
out $\varphi $ we find

\begin{equation}
\varphi =(\frac{\tilde{g}^{AB}h_{AB}}{p+1})^{\frac{p-1}{2}}.  \eqnum{76}
\end{equation}
Substituting this result for $\varphi $ into (75) we get

\begin{equation}
S=\frac 1{(p+1)^{\frac{p+1}2}}\int \frac{d^{p+1}y}{T_p}\sqrt{-\tilde{g}}(%
\tilde{g}^{AB}h_{AB})^{\frac{p+1}2}  \eqnum{77}
\end{equation}
which is the Weyl invariant action for p-branes. Therefore, we have shown
that the Weyl invariant action (77) follows from the gravity-dilaton action
(55) when the constants $\Lambda $ and $a$ have the form (72) and (73)
respectively, that is, we have shown a mechanism to derive (74) from the
following action

\begin{equation}
S=-%
%TCIMACRO{\tfrac{1}{16\pi G_{D+1}} }%
%BeginExpansion
{\textstyle{1 \over 16\pi G_{D+1}}}%
%EndExpansion
\int d^{D+1}y\sqrt{-g}e^{-\phi }(\varphi (R+(\nabla \phi )^{2})+(p-1)\varphi
^{\frac{p+1}{p-1}}),  \eqnum{78}
\end{equation}

Integrating out $\varphi $ in (78) we derive the action

\begin{equation}
S=-%
%TCIMACRO{\tfrac{1}{16\pi G_{D+1}(p+1)^{\frac{p+1}{2}}} }%
%BeginExpansion
{\textstyle{1 \over 16\pi G_{D+1}(p+1)^{\frac{p+1}{2}}}}%
%EndExpansion
\int d^{D+1}y\sqrt{-g}e^{-\phi }(R+(\nabla \phi )^{2})^{\frac{p+1}{2}}, 
\eqnum{81}
\end{equation}
Therefore, we have a higher curvature theory as an starting point. It is
interesting to note that only for the case $p=1,$ corresponding to strings,
(81) turns out to be an action linear in curvature tensor $R$.

\smallskip \smallskip\ 

\noindent {\bf 6.- FINAL COMMENTS}

\bigskip

In this article we have clarified some aspects of the Weyl invariant actions
for p-branes, Dp-branes and null p-branes. In sections 2, 3 and 4 we
develope a mechanism to derive different Weyl invariant p-branes and
Dp-branes actions. While in section 5 we have shown a geometric origin of
Weyl invarint actions for p-branes.

So far, our treatment of the equivalence of various Weyl invariant actions
has been classical, without taking into account certain determinants of the
associated partition function. Such determinants may lead to some anomalies
allowing equivalence only under certain conditions. For instance, at the
quantum level the Polyakov type action for p-branes (7) is equivalent to the
Dirac-Nambu-Goto type action (1) only if we set $d=25.$

The supersymmetrization of the procedure proposed in second, third and
fourth sections seem to be straightforward. One needs simply to use the
prescription

\begin{equation}
\partial _{A}x^{\hat{\mu}}\rightarrow E_{A}^{\hat{\mu}}=\partial _{A}Z^{\hat{%
M}}\Pi _{\hat{M}}^{\hat{\mu}},  \eqnum{82}
\end{equation}
where $Z^{\hat{M}}=(x^{\hat{\mu}},\theta ^{\hat{\alpha}})$ are the
coordinates of a superspace and $\Pi _{\hat{M}}^{\hat{\mu}}$ are the
supervielbein. In this case the action (13) can be generalized to

\begin{equation}
\begin{array}{c}
S_{p}^{W_{1}}=-\frac{T_{p}}{2}\int d^{p+1}\xi \sqrt{-g}(\Phi ^{\frac{p-1}{2}%
}g^{AB}E_{A}^{\hat{\mu}}E_{B}^{\hat{\nu}}\eta _{\hat{\mu}\hat{\nu}%
}-(p-1)\Phi ^{\frac{p+1}{2}}) \\ 
\\ 
+\frac{1}{(p+1)!}\varepsilon ^{A_{1}...A_{p+1}}E_{A_{1}}^{\hat{\mu}%
_{1}}...E_{A_{p+1}}^{\hat{\mu}_{p+1}}A_{\hat{\mu}_{1}}..._{\hat{\mu}_{p+1}},
\end{array}
\eqnum{83}
\end{equation}
where $A_{\hat{\mu}_{1}}..._{\hat{\mu}_{p+1}}$ is a completely antisymmetric
gauge field tensor. This action is invariant under the world volume
diffeomorphisms and the Weyl tranformation (14)-(15).

It may be interesting for further works to follow similar procedure as the
one of section 5 to find the super p-branes actions from a higher
dimensional supergravity action. For instance, this may help to understand
the the so called $\kappa $ symmetry from a geometrical perspective.

Recently, a Weyl invariant spinning membrane action [18] was constructed
where the conformal symmetry and S-supersymmetry are broken. It may be
interesting to relate the present work with such a system.

\begin{center}
{\bf Acknowledgments}
\end{center}

We are indebted with I. Galaviz for helpful discussions.

\bigskip

\smallskip\ \

\end{document}